\documentclass[a4paper]{article}
\usepackage[margin=3cm]{geometry}
\usepackage{latexsym}
\usepackage{amsmath}
\usepackage{amsthm}
\usepackage{amssymb}
\usepackage{amsfonts}
\usepackage{graphicx}
\usepackage{cite}

\makeatletter
\def\@biblabel#1{}
\makeatother

\begin{document}

\title{BREAKING POINTS IN QUARTIC MAPS}
\author{M. ROMERA, G. PASTOR, A. MARTIN, A.B. ORUE, F. MONTOYA\\
Instituto de Tecnolog\'{\i}as F\'{\i}sicas y de la Informaci\'{o}n (ITEFI)\\Consejo Superior de Investigaciones Cient\'{\i}ficas (CSIC)\\ Serrano 144, Madrid, 28006, Spain\\miguel.romera.garcia@gmail.com\\\{gerardo, agustin, amalia.orue, fausto\}@iec.csic.es\\
M.-F. DANCA\\
Dept. of Mathematics and Computer Science, ``Avram Iancu'' University, Ilie M\u{a}celaru 1A\\ Cluj-Napoca, 400380, Rom\^{a}nia\\
Also at Romanian Institute of Science and Technology,\\ Cire\c{s}ilor 29, Cluj-Napoca, 400487, Rom\^{a}nia.\\
danca@rist.ro}

\maketitle

\begin{abstract}
\noindent Dynamical systems, whether continuous or discrete, are used by physicists in order to study non-linear phenomena. In the case of discrete dynamical systems, one of the most used is the quadratic map depending on a parameter. However, some phenomena can depend alternatively of two values of the same parameter. We use the quadratic map $x_{n+1} =1-ax_{n}^{2} $ when the parameter alternates between two values during the iteration process. In this case, the orbit of the alternate system is the sum of the orbits of two quartic maps. The bifurcation diagrams of these maps present breaking points where abruptly change their evolution.
\end{abstract}


\section{\label{sec:intro}Introduction}
\noindent In ecological modeling, seasonality can be represented as a switching between different environmental conditions in discrete dynamical systems. Maier and Peacock-L\'{o}pez~[2010] iterated the quadratic logistic map using this switching strategy, i.e., with one pa\-ra\-me\-ter value when iteration is odd and another value when the iteration is even. The bifurcation diagram of Fig. 2 of such a paper~\cite{Maier2010}, obtained with a fixed value of the odd parameter and a variable value of the even parameter, presents two features not mentioned by the authors: 1) it is double (the meaning of double will be explained in paragraph~\ref{ssec:bifurcation-diagrams}), and 2) it exhibits points where abruptly changes its evolution. These features will be analyzed in this paper by using the alternate iteration of the quadratic map $1-ax^{2} $.

 The alternating system $x_{n+1} =\left\{\begin{array}{c} {f(x_{n} ){\rm \; if\; }n{\rm \; is\; even}} \\ {g(x_{n}){\rm \; if\; }n{\rm \; is\; odd}} \end{array}\right. $ also has been studied from a general point of view in~\cite{DAnielloSteele2011,DAnielloOliveira2009}. AlSharawi and Angelos~[2006] studied the p-periodic logistic equation, alternating periodically the parameter of the map. The bifurcation diagram of Fig.~2 in \cite{AlSharawiAngelos2006}, for the 2-periodic logistic equation, is double although it can not show points where its evolution changes abruptly because of the parametric representation of this figure.

J\'{a}nosi and Gallas~[1999] studied the ``restricted'' quartic map $1-a(1-ax^{2})^{2}$, the second iteration of the map $1-ax^{2} $, where the bifurcation diagram presented an  ``explosion'' of the chaotic amplitude when $a\approx 1.54365$ (see Fig.~1 in~\cite{JGallas1999}). As we will see in Section~\ref{ss:break-points-insts-case2} this explosion is a breaking point. These authors also formulated the ``generic'' quartic map $1-a(1-bx^{2})^{2} $, that we study in this paper (where $b=a_{*}$), but it is not studied in~\cite{JGallas1999}. Gallas studied the quartic map $(a-x_{}^{2} )^{2} -b$ in~\cite{Gallas1993,Gallas1994,Gallas1995}, but this map has different dynamics because it is not topologically conjugate with $1-a(1-bx^{2} )^{2} $ according to~\cite{GrossmannThomae1997,MilnorThurston1988}.

 Some of us~\cite{DancaRomera2009} have presented the alternate Julia sets obtained by alternate iteration of two maps $z_{n+1}~=~z_{n}^{2}~+~c_{i},{\rm \;}i=1,2$ and proved that these sets can be connected, disconnected or totally disconnected verifying the known Fatou-Julia theorem in the case of polynomials of degree greater than two.

 In this paper we use the alternate iteration of the quadratic map $1-ax^{2} $, where the parameter takes the values $a$ and $a_{*} $, to study the bifurcation diagram of a 2-periodic quadratic system (the iteration with the quadratic map $1-ax^{2} $, instead the logistic map, presents notation advantages). In Section~\ref{sec:alternate-quartic} we show that the orbit of the alternate iteration is the sum of the orbits of the quartic maps $x_{n+1} =1-a(1-a_{*} x_{n}^{2} )^{2} $ and ${\rm \; }x_{n+1} =1-a_{*} (1-ax_{n}^{2} )^{2} $. In Section~\ref{sec:Breaking-points} we show that the bifurcation diagrams of these quartic maps exhibit breaking points, and we obtain formulas to calculate these points. Finally, examples and conclusions are shown in Sections~\ref{sec:examples} and~\ref{sec:conclusion} respectively.

\section{Alternate Quadratic System and Quartic Maps}\label{sec:alternate-quartic}
Let us consider the quadratic map~\cite{PostCapel1991},
 \begin{equation}\label{ec:qmap}
    x_{n+1} =1-ax_{n}^{2}
 \end{equation}
when the parameter takes alternatively the values $a$ and $a_{*} $. More precisely, let us consider the systems,
\begin{equation}
  A:\; \; x_{n+1}=\biggl\{
  \begin{aligned}
  &1-ax_{n}^{2} \quad\text{if}& n \text{ is\; even}  \\
  &1-a_{*} x_{n}^{2}\quad\text{if}& n \text{ is\; odd}
  \end{aligned}
\end{equation}
and
  \begin{equation}
  B:\; \; x_{n+1}=\biggl\{
  \begin{aligned}
  &1-a_{*} x_{n}^{2} \quad\text{if}& n \text{ is\; even}  \\
  &1-ax_{n}^{2} \quad\text{if}& n \text{ is\; odd.}
  \end{aligned}
\end{equation}
 Starting from the critical point 0, we have,
 \begin{eqnarray*}
   x_{0}^{{\rm A}} &=& 0, \\
   x_{1}^{{\rm A}} &=& 1, \\
   x_{2}^{{\rm A}} &=& 1-a_{*}, \\
   x_{3}^{{\rm A}} &=& 1-a(1-a_{*} )^{2}, \\
   x_{4}^{{\rm A}} &=& 1-a_{*} [1-a(1-a_{*} )^{2} ]^{2},\\
   x_{5}^{{\rm A}} &=& 1-a\left\{1-a_{*} [1-a(1-a_{*} )^{2} ]^{2} \right\}^{2}\ldots
 \end{eqnarray*}
and
\begin{eqnarray*}
   x_{0}^{{\rm B}} &=& 0, \\
   x_{1}^{{\rm B}} &=& 1, \\
   x_{2}^{{\rm B}} &=& 1-a, \\
   x_{3}^{{\rm B}} &=& 1-a_{*} (1-a)^{2}, \\
   x_{4}^{{\rm B}} &=& 1-a[1-a_{*} (1-a)^{2} ]^{2},\\
   x_{5}^{{\rm B}} &=& 1-a_{*} \left\{1-a[1-a_{*} (1-a)^{2} ]^{2} \right\}^{2}\ldots
 \end{eqnarray*}

It is easy verify the following

\vspace{35mm}
\emph{Property} 1. Let us consider the quartics ${\rm C:\; }x_{n+1} =1-a(1-a_{*} x_{n}^{2} )^{2} $ and ${\rm D:\; }x_{n+1} =1-a_{*} (1-ax_{n}^{2} )^{2} $. The orbit of the critical point $0$ of the alternate quadratic system A(B) can be obtained by superposing the orbit of the critical value 1 of the quartic C(D) with the orbit of the critical point $0$ of the quartic D(C).

\section{Breaking Points in Quartic Maps}\label{sec:Breaking-points}
\begin{figure} [b]
\begin{center}
\includegraphics[width=8.5cm]{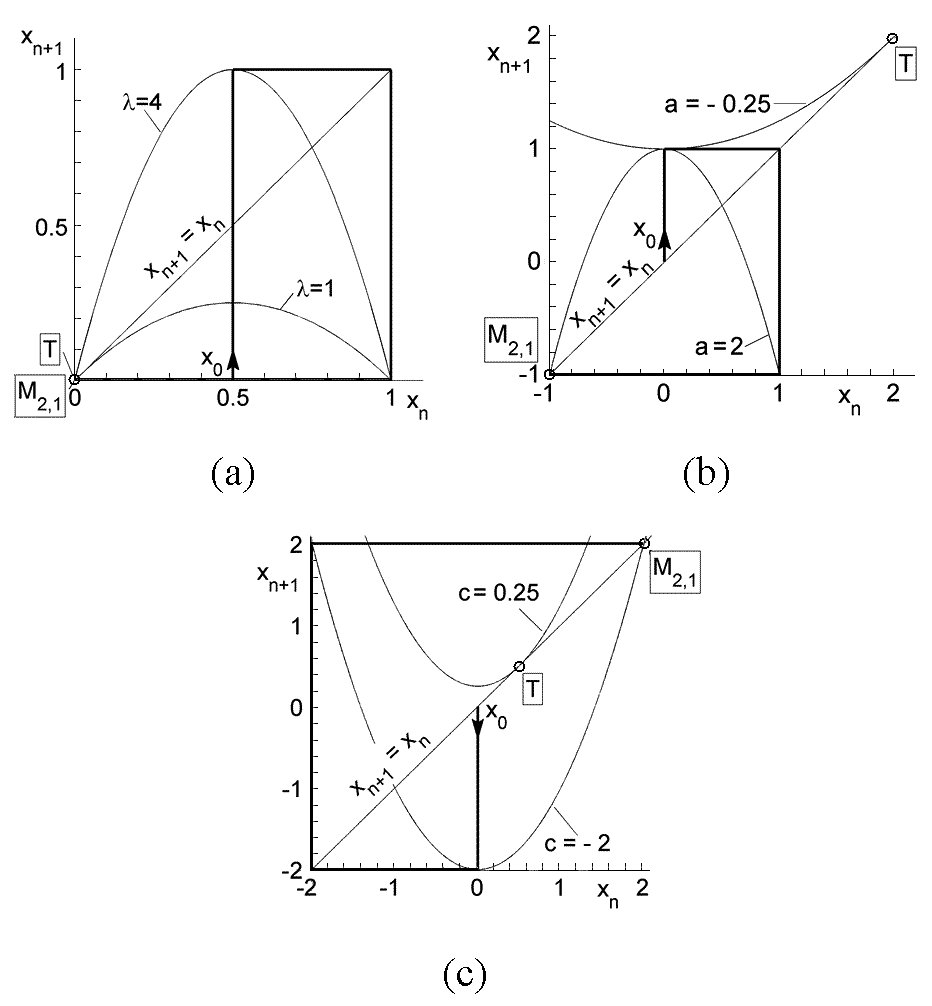}\\
\caption{Location of points T and $\textrm{M}_{2,1}$ in quadratic maps. (a) $x_{n+1}=\lambda x_n(1-x_n)$. (b) $x_{n+1}=1-a x_n^2$. (c) $x_{n+1}= x_n^2+c.$ } \label{fig:1}
\end{center}
\end{figure}
\begin{figure} [b]
\begin{center}
\includegraphics[width=8.5cm]{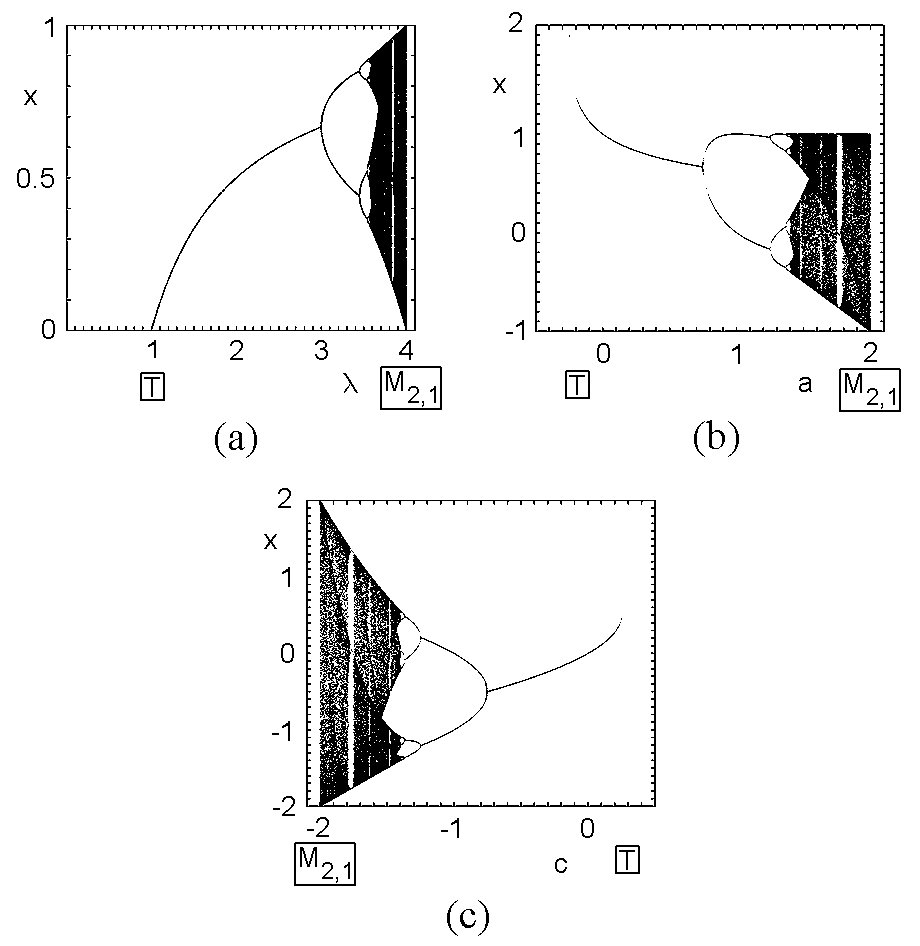}\\
\caption{Points T and $\textrm{M}_{2,1}$ in the bifurcation diagrams of quadratic maps. (a) $x_{n+1}=\lambda x_n(1-x_n)$. (b) $x_{n+1}=1-a x_n^2$. (c) $x_{n+1}= x_n^2+c$.} \label{fig:2}
\end{center}
\end{figure}

\noindent Let us briefly recall the quadratic maps. Figure~\ref{fig:1} shows the graphs of the maps $x_{n+1} =\lambda x_{n} (1-x_{n} )$~\cite{May1976}, $x_{n+1} =1-ax_{n}^{2}$~\cite{PostCapel1991}, $x_{n+1} =x_{n}^{2} +c$~\cite{PastorRomeraMontoya1996} and the graphical iteration starting from their critical points in two cases. Firstly, for the parameter values when the graphs are tangent at points T to the straight line $x_{n+1} =x_{n} $. Secondly, for the parameter values when the graphical iterations go to Misiurewicz points ${\rm M}_{2,1} $ where the orbit is preperiodic (the first subscript is the preperiod and the second one is the period) \cite{Misiurewicz1991}. Points T occur when $\lambda =1$, $a=-0.25$ and $c=0.25$. Points ${\rm M}_{2,1} $ occur when $\lambda =4$, $a=2$ and $c=-2$. Note that a quadratic map only has one pa\-ra\-me\-ter value causing a point T and one parameter value causing a point ${\rm M}_{2,1} $. The points T and ${\rm M}_{2,1} $ are the beginning and the end of the bifurcation diagrams of these quadratic maps, see Fig.~\ref{fig:2}.
Other tangent point T and Misiurewicz point M (with low preperiod and period) can appear inside the bifurcation diagram of a quartic map. When this occurs, the bifurcation diagram shows a breaking point by tangency (new T) or a breaking point by instability (new M).

\subsection{Breaking point by tangency}
Let us consider the family of quartics $${\rm C}:{\rm \; }x_{n+1} =1-a(1-a_{*} x_{n}^{2} )^{2}$$ when $a_{*} ={\rm constant}$ and the parameter of the family is $a$. We are interested in knowing if two maps of this family can have their graphs tangent to the straight line $x_{n+1} =x_{n} $ for two values of the parameter. For this, the first derivative in the tangent points must be equal to unity, i.e.
\begin{equation}\label{ec:BPbytangency}
    4aa_{*} x(1-a_{*} x_{}^{{\rm 2}})=1.
\end{equation}

The tangent points must belong to $x_{n} =x_{n+1} $, i.e., $1-a(1-a_{*} x^{2} )^{2} =x$ that can be written as
\begin{equation}\label{ec:BPbytangency2}
    a(1-a_{*} x^{2} )^{2} =1-x.
\end{equation}

From Eq.~\eqref{ec:BPbytangency} and Eq.~\eqref{ec:BPbytangency2} we obtain ${(1-a_{*} x^{2} )\mathord{\left/ {\vphantom {(1-a_{*} x^{2} ) 4a_{*} x}} \right. \kern-\nulldelimiterspace} 4a_{*} x} =1-x$ and the tangent points,
\begin{equation}\label{ec:tangentpoint}
    {x_{{\rm TC}} =[2a_{*} \pm (4a_{*}^{2} -3a_{*} )^{1/2} ]\mathord{\left/ {\vphantom {x_{{\rm TC}} =[2a_{*} \pm (4a_{*}^{2} -3a_{*} )^{1/2} ] 3a_{*} }} \right. \kern-\nulldelimiterspace} 3a_{*}}. \end{equation}

According to Eq.~\eqref{ec:tangentpoint}, when $a_{*} >{3\mathord{\left/ {\vphantom {3 4}} \right. \kern-\nulldelimiterspace} 4} $ there exist two tangent points of quartic C with the straight line $x_{n+1} =x_{n} $ whose parameter values are given by
\begin{equation}\label{ec:2tangentpointparam}
    a={1\mathord{\left/ {\vphantom {1 4a_{*} x_{{\rm TC}} (1-a_{*} x_{{\rm TC}}^{{\rm 2}} )}} \right. \kern-\nulldelimiterspace} 4a_{*} x_{{\rm TC}} (1-a_{*} x_{{\rm TC}}^{{\rm 2}} )} , \end{equation}
where one of these tangent points corresponds to the beginning of the bifurcation diagram of the quartic C and the other is a breaking point by tangency.

Similarly, starting from the family of quartics ${\rm D}:{\rm \; }1-a_{*} (1-ax^{2} )^{2} $, when $a={\rm constant}$ and the parameter is $a_{*} $ we obtain
\begin{equation}\label{ec:quarticsD1}
    x_{{\rm TD}} ={[2a\pm (4a_{}^{2} -3a)^{1/2} ]\mathord{\left/ {\vphantom {[2a\pm (4a_{}^{2} -3a)^{1/2} ] 3a}} \right. \kern-\nulldelimiterspace} 3a}
\end{equation}
and
\begin{equation}\label{ec:quarticsD2}
    a_{*} ={1\mathord{\left/ {\vphantom {1 4ax_{{\rm TD}} (1-ax_{{\rm TD}}^{{\rm 2}} )}} \right. \kern-\nulldelimiterspace} 4ax_{{\rm TD}} (1-ax_{{\rm TD}}^{{\rm 2}} )} .
\end{equation}

It is easy to see that if the quartic C is tangent to the straight line $x_{n} =x_{n+1} $ for a given pair $(a,a_{*})$, then the quartic D is also tangent to $x_{n} =x_{n+1} $ for the same pair of parameter values.

\subsection{Breaking points by instability}

Let us consider the quartic C: $f_{{\rm C}} (x)=1-a(1-a_{*} x^{2} )^{2}$ with the critical points $0$ and $\pm ({1\mathord{\left/ {\vphantom {1 a_{*} }} \right. \kern-\nulldelimiterspace} a_{*} } )^{1/2} $. The critical values are $1-a$ and 1.
The orbit of the critical point $0$ is $$0,\; 1-a,\; f_{{\rm C}} (1-a),\; f_{{\rm C}} [f_{{\rm C}} (1-a)]\dots$$
and the orbit of the critical value 1 is $$ 1,\; f_{{\rm C}} (1), \;f_{{\rm C}} [f_{{\rm C}} (1)]\dots.$$ We will study the following cases:

\subsubsection{Case: $1-a=f_{{\rm C}} (1-a)$}

If $1-a=f_{{\rm C}} (1-a)$ the orbit of the critical point $0$ is: $0,\; 1-a,\; 1-a,\; 1-a,\dots$ It has preperiod-1 and period-1, i.e., the parameter value is a Misiurewicz point ${\rm M}_{1,1} $. The equation $1-a=f_{{\rm C}} (1-a)$ has five solutions for $a$ when $a_{*} ={\rm constant}$. Ignoring the trivial solutions $a=0$ and $a=1$ (double), we obtain the remaining two solutions by means of the equation
\begin{equation}\label{ec:solcase1-1}
    a=1\pm ({2\mathord{\left/ {\vphantom {2 a_{*} }} \right. \kern-\nulldelimiterspace} a_{*} } )^{1/2}.
\end{equation}

When $a={\rm constant}$, the equation $1-a=f_{{\rm C}} (1-a)$ has two solutions. Ignoring the trivial solution $a_{*} =0$, we have
 \begin{equation}\label{ec:solcase1-2}
    a_{*} ={2\mathord{\left/ {\vphantom {2 (1-a)^{2} }} \right. \kern-\nulldelimiterspace} (1-a)^{2}}.
\end{equation}

We must verify a posteriori if the parameter values obtained by Eq.~\eqref{ec:solcase1-1} and Eq.~\eqref{ec:solcase1-2} give rise to Misiurewicz points, i.e., if the slope of the graph in these points is greater than unity. When this occurs they will be breaking points if, besides, they are inside the working interval of the bifurcation diagram.

\subsubsection{Case: $1-a\ne f_{{\rm C}} (1-a)=f_{{\rm C}} [f_{{\rm C}} (1-a)]$}\label{ss:break-points-insts-case2}
If $1-a\ne f_{{\rm C}} (1-a)=f_{{\rm C}} [f_{{\rm C}} (1-a)]$ the orbit of the critical point $0$ is: $0,\; 1-a,\; f_{{\rm C}} (1-a),\; f_{{\rm C}} (1-a),\dots$ It has preperiod-2 and period-1, i.e., the parameter value is a Misiurewicz point ${\rm M}_{{\rm 2,1}} $. As is easy to see by graphical iteration, the breaking point occurs when ${\rm M}_{{\rm 2,1}} $ is between the critical points $0$ and $({1\mathord{\left/ {\vphantom {1 a_{*} }} \right. \kern-\nulldelimiterspace} a_{*} } )^{1/2} $, and the first iterate is on the right of the critical point $-({1\mathord{\left/ {\vphantom {1 a_{*} }} \right. \kern-\nulldelimiterspace} a_{*} } )^{1/2} $, i.e., if  it is satisfied that
\begin{equation}
  \biggl\{ \begin{aligned}\label{ec:case2conditions}
  &0<f_{{\rm C}} (1-a)< ({1\mathord{\left/ {\vphantom {1 a_{*} }}\right.\kern-\nulldelimiterspace   } a_{*} } )^{1/2} \\
  &1-a> -(1/a_{*} )^{1/2}.
  \end{aligned}
\end{equation}

When $a_{*} ={\rm constant}$, the equation $f_{{\rm C}} (1-a)=f_{{\rm C}} [f_{{\rm C}} (1-a)]$ has twenty one solutions for $a$. Ignoring the trivial solutions $a=0$ (double), $a=1$ (triple), and the two solutions of Eq.~\eqref{ec:solcase1-1}, we obtain the remaining fourteen solutions for $a$ by means of the equations
\begin{equation}\label{ec:4solutionfora}
  \begin{aligned}
  & a_{*}^{2} a_{}^{4} -3a_{*}^{2} a_{}^{3} +(3a_{*}^{2} -2a_{*} )a_{}^{2} +(2a_{*} -a_{*}^{2} )a_{} +2=0\\
  &\text{(four solutions for $a$)}
  \end{aligned}
\end{equation}
and
\begin{equation}\label{ec:tensolutionfora}
  \begin{aligned}
  & 2-a_{*} (1-a)^{2} -a_{*} \left\{1-a[1-a_{*} (1-a)^{2} ]^{2} \right\}^{2} =0\\
  &\text{(ten solutions for $a$)}.
  \end{aligned}
\end{equation}

\noindent We can find numerically the real solutions of Eq.~\eqref{ec:4solutionfora} and Eq.~\eqref{ec:tensolutionfora} and check whether these solutions verify Eq.~\eqref{ec:case2conditions} to see if they are breaking points.
\begin{figure*}
\begin{center}
\includegraphics[width=165mm]{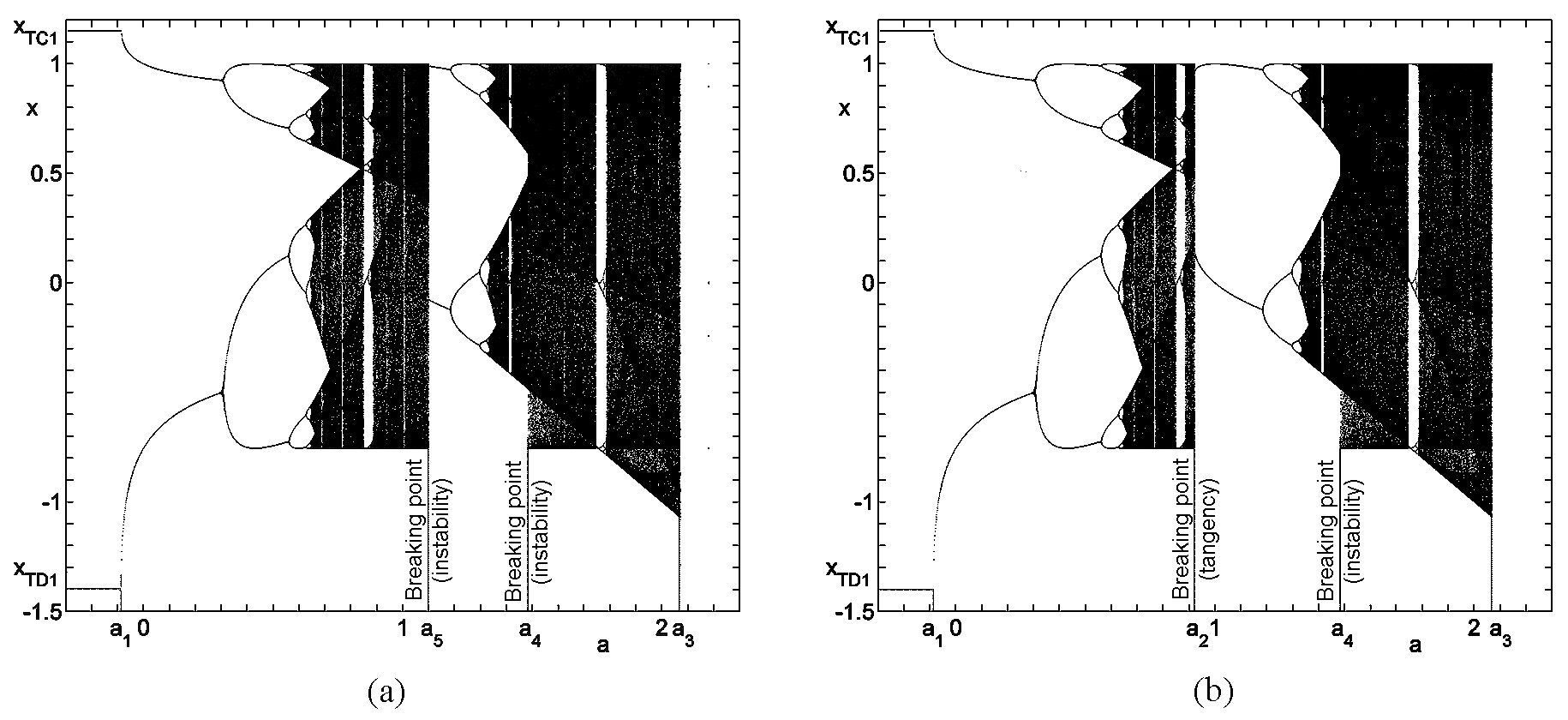}\\
\caption{Bifurcation diagrams of the alternate quadratic systems A and B when  $a_*=1.754877$.  (a) System A. (b) System B.} \label{fig:3}
\end{center}
\end{figure*}

When $a={\rm constant}$, the equation $f_{{\rm C}} (1-a)=f_{{\rm C}} [f_{{\rm C}} (1-a)]$ has ten solutions for $a_{*} $. Ignoring the trivial solution $a_{*} =0$ (double) and the solution of Eq.~\eqref{ec:solcase1-2}, we obtain the remaining seven solutions for $a_{*} $ by means of
\begin{equation}\label{ec:2solutionforaasterisco}
  \begin{aligned}
  & a_{*} ={[1\pm ({-1+2\mathord{\left/ {\vphantom {-1+2 a}} \right. \kern-\nulldelimiterspace} a} )^{1/2} ]\mathord{\left/ {\vphantom {[1\pm ({-1+2\mathord{\left/ {\vphantom {-1+2 a}} \right. \kern-\nulldelimiterspace} a} )^{1/2} ] (1-a)^{2} }} \right. \kern-\nulldelimiterspace} (1-a)^{2} }\\
  &\text{(two solutions for $a_{*} $)}
  \end{aligned}
\end{equation}
and Eq.~\eqref{ec:tensolutionfora} (five solutions for $a_{*} $).

 The ``explosion'' of the chaotic amplitude observed in the map $f(x)=1-a(1-ax^{2} )^{2} $ when $a\approx 1.54365$~\cite{JGallas1999} is a breaking point by instability due to a Misiurewicz point ${\rm M}_{{\rm 2,1}} $. We have
\begin{eqnarray*}
   x_{0} &=& 0, \\
   x_{1} &=& 1-a\approx -0.54365, \\
   x_{2} &=& f(1-a)\approx 0.54357, \\
   x_{3} &=& f[f(1-a)]\approx 0.54334\approx x_{2}.
  \end{eqnarray*}
\subsubsection{Case: $f_{{\rm C}} (1)=f_{{\rm C}} [f_{{\rm C}} (1)]$}

If $f_{{\rm C}} (1)=f_{{\rm C}} [f_{{\rm C}} (1)]$ the orbit of the critical value $1$ is: $1,\; f_{{\rm C}} (1),\; f_{{\rm C}} (1),\dots$ It has preperiod-1 and period-1, i.e., the parameter value is a Misiurewicz point  ${\rm M}_{{\rm 1,1}} $. When $a_{*} ={\rm constant}$, the equation $f_{{\rm C}} (1)=f_{{\rm C}} [f_{{\rm C}} (1)]$ has five solutions. Ignoring the trivial solution $a=0$ (double) we have
\begin{equation}\label{ec:case3-1}
    a=2/(1-a_*)^2
\end{equation}
and
\begin{equation}\label{ec:case3-2}
    a=[1\pm (-1+2/a_*)^{1/2}] \Big{/}(1-a_*)^2.
\end{equation}

Note that Eq.~\eqref{ec:case3-1} and Eq.~\eqref{ec:case3-2} are Eq.~\eqref{ec:solcase1-2} and Eq.~\eqref{ec:2solutionforaasterisco} with $a$ and $a_{*} $ interchanged.

When $a={\rm constant}$, the equation $f_{{\rm C}} (1)=f_{{\rm C}} [f_{{\rm C}} (1)]$ has ten solutions. Ignoring the trivial solutions $a_{*} =0$ and $a_{*} =1$ (triple), we have possible breaking points at the parameter values given by
\begin{equation}\label{ec:case3posibleBP}
    a_{*} =1\pm ({2\mathord{\left/ {\vphantom {2 a}} \right. \kern-\nulldelimiterspace} a} )^{1/2}
\end{equation}
and the real solutions of
\begin{equation}\label{ec:case3realsolutions}
    a_{}^{2} a_{*}^{4} -3a_{}^{2} a_{*}^{3} +(3a_{}^{2} -2a)a_{*}^{2} +(2a-a_{}^{2} )a_{*} +2=0.
 \end{equation}

Note that Eq.~\eqref{ec:case3posibleBP} and Eq.~\eqref{ec:case3realsolutions} are  Eq.~\eqref{ec:solcase1-1} and Eq.~\eqref{ec:4solutionfora} with $a$ and $a_{*} $ interchanged.

As we can see by graphical iteration, the breaking point occurs when ${\rm M}_{{\rm 1,1}} $ is between the critical points $0$ and $({1\mathord{\left/ {\vphantom {1 a_{*} }} \right. \kern-\nulldelimiterspace} a_{*} } )^{1/2} $, i.e., if
\begin{equation}\label{ec:case3BPcondition}
    0<f_{{\rm C}} (1)<({1\mathord{\left/ {\vphantom {1 a_{*} }} \right. \kern-\nulldelimiterspace} a_{*} } )^{1/2}.
\end{equation}

\section{Examples}\label{sec:examples}

\subsection{Bifurcation diagrams when $a_{*}=1.754877$}\label{ssec:bifurcation-diagrams}

\begin{figure*}[thbp]
\begin{center}
\includegraphics[width=165mm]{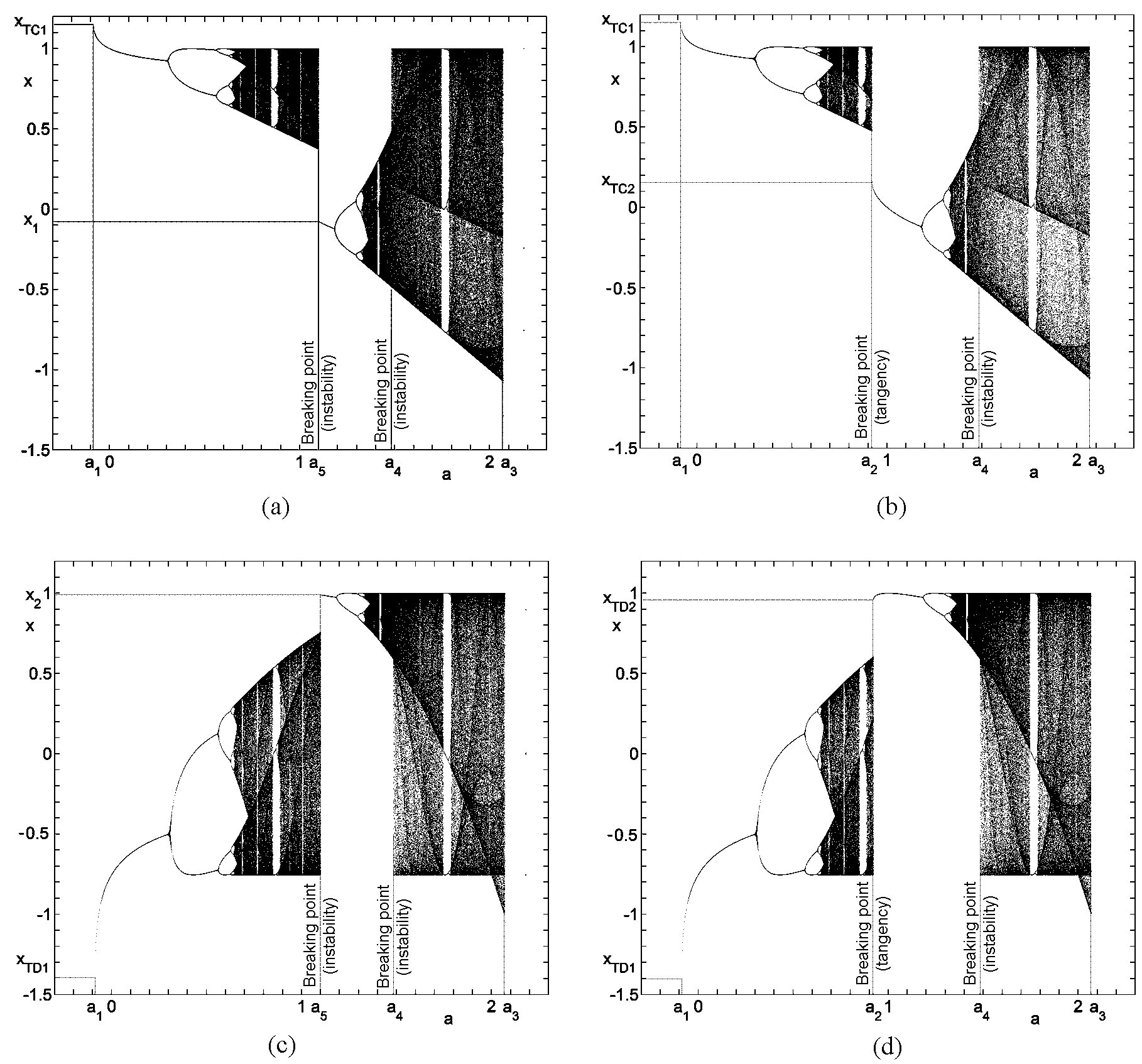}\\
\caption{Bifurcation diagrams of quartics C: $x_{n+1}=1-a(1-a_{*}x_n^2)^2$  and D: $x_{n+1}=1-a_*(1-ax_n^2)^2$  when  $a_*=1.754877$. (a) Quartic C, with  $x_0=1$. (b) Quartic C, with  $x_0=0$. (c) Quartic D, with  $x_0=0$. (d) Quartic D, with  $x_0=1$.} \label{fig:4}
\end{center}
\end{figure*}

 The systems A and B have, obviously, two parameters and to obtain the bifurcation diagram of these systems it is necessary to fix one of them. Taking into account that the map $1-ax^{2} $ has a superstable period-3 orbit when $a=1.754877$ (see the period-3 window at this parameter value in Fig.~\ref{fig:2}(b)) let us consider, as an example, the bifurcation diagrams of systems A and B (Fig.~\ref{fig:3}) and quartics C and D (Fig.~\ref{fig:4}) when $a_{*} ={\rm 1.754877}$ and the parameter is $a$. As a consequence of Property 1, the bifurcation diagrams of the alternate quadratic systems A and B are double in the sense that each one of them is constituted by the bifurcation diagrams of two quartic maps that operate simultaneously. So, the bifurcation diagram of the alternate quadratic system A, Fig.~\ref{fig:3}(a), is the bifurcation diagram of the quartic C with $x_0=1$, Fig.~\ref{fig:4}(a), plus the bifurcation diagram of the quartic D with $x_0=0$, Fig.~\ref{fig:4}(c). Analogously, the bifurcation diagram of the alternate quadratic system B, Fig.~\ref{fig:3}(b), is the bifurcation diagram of quartic C with $x_0=0$, Fig.~\ref{fig:4}(b), plus the bifurcation diagram of quartic D with $x_0=1$,  Fig.~\ref{fig:4}(d).

\subsubsection{Breaking points by tangency}

Equation~\eqref{ec:tangentpoint} gives the tangent points of quartic C, $x_{{\rm TC1}} =1.171144$ and $x_{{\rm TC2}} =0.162189$, and Eq.~\eqref{ec:2tangentpointparam} gives the parameter values of the quartic C, $a_{1} =\!-0.086457$ and $a_{2} =0.920867$.

\begin{figure} [thbp]
\begin{center}
\includegraphics[width=8.5cm]{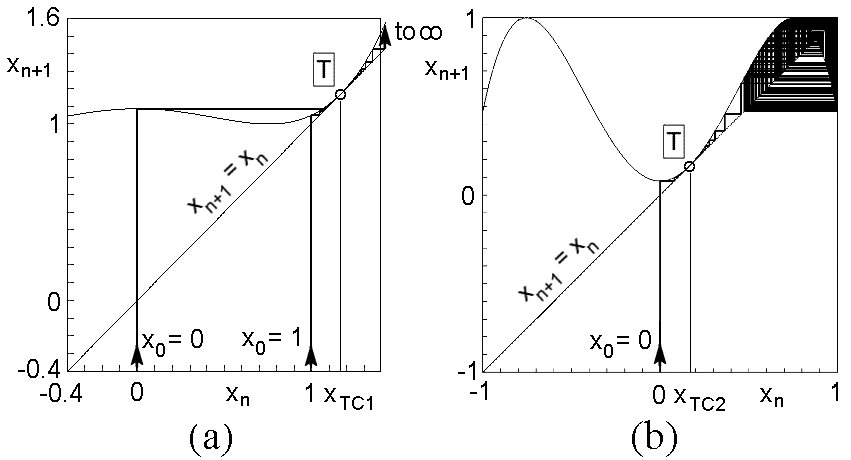}\\
\caption{Tangent points of quartic $x_{n+1}=1-a(1-a_{*}x_n^2)^2$ with $a_*=1.754877$. (a) $a=a_1= -0.086457$. (b) $a=a_2=0.920867$.} \label{fig:5}
\end{center}
\end{figure}

If $a=a_{1} $, the orbit reaches the tangent point $x_{{\rm TC1}} $ and if $a<a_{1} $ the orbit escapes to infinity, Fig.~\ref{fig:5}(a). Equation~\eqref{ec:quarticsD1} gives two values of $x_{{\rm TD}} $ for quartic D, but only $x_{{\rm TD1}} =\!-1.406959$ is a valid solution because Eq.~\eqref{ec:quarticsD2} returns $a_{*} =1.754877$, Fig.~\ref{fig:3} and Fig.~\ref{fig:4}(c,d). The parameter value $a_{1} $ corresponds to the lower extreme of the bifurcation diagram of systems A and B and it is not a breaking point.\vskip5pt

The parameter value $a_{2} $ is inside the bifurcation diagram, when the orbit reaches the tangent point $x_{{\rm TC2}} $, Fig.~\ref{fig:5}(b). As we can see in Fig.~\ref{fig:3}(b) and Fig.~\ref{fig:4}(b,d), if $a$ is slightly smaller than $a_{2} $, the orbit is chaotic; if $a$ is slightly greater than $a_{2} $, the orbit has period-2 in the alternate system B, period-1 in quartic C and period-1 in quartic D. When $a=a_{2} $, Eq.~\eqref{ec:quarticsD1} gives the valid solution $x_{{\rm TD2}} =0.953836$ because Eq.~\eqref{ec:quarticsD2} returns $a_{*} =1.754877$; the other solution is not valid. The parameter value $a_{2} $ is a breaking point.

\subsubsection{Breaking points by instability}
\begin{figure} [thbp]
\begin{center}
\includegraphics[width=8.5cm]{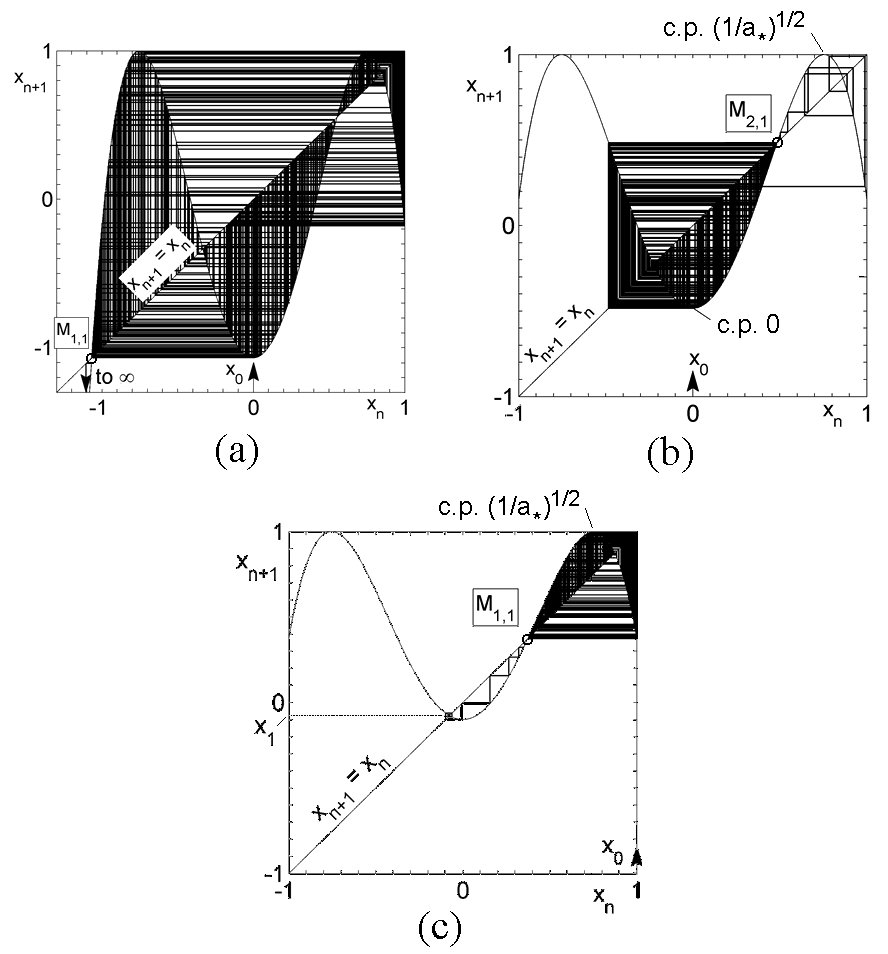}\\
\caption{Misiurewicz points in quartic $x_{n+1}=1-a(1-a_{*}x_n^2)^2$ with $a_*=1.754877$. (a) $a=a_3=2.067558$ and $x_0=0$. (b) $a=a_4=1.483181$ and $x_0=0$. (c) $a=a_5=1.099012$ and $x_0=1$.} \label{fig:6}
\end{center}
\end{figure}

Equation~\eqref{ec:solcase1-1} has two solutions. The solution $a_{3} =2.067558$ gives rise to a Misiurewicz point ${\rm M}_{{\rm 1,1}} $, Fig.~\ref{fig:6}(a). If $a<a_{3} $ the orbit is chaotic, and if $a>a_{3} $ the orbit escapes to infinity. The parameter value $a_{3} $ is the upper extreme of the working interval of both systems A and B. For this reason, it is not a breaking point. The other solution is not a breaking point because the slope of the graph in the point ${\rm M}_{{\rm 1,1}} $ is less than unity.

Equation~\eqref{ec:4solutionfora} has two real solutions. The solution $a_{4} =1.483181$ originates a Misiurewicz point ${\rm M}_{{\rm 2,1}} $ verifying the conditions for breaking point of Eq.~\eqref{ec:case2conditions}, Fig.~\ref{fig:6}(b). If $a<a_{4} $, the iterations always fall inside the attraction basin of the critical point $0$. However, if $a>a_{4} $ the iterations fall inside the attraction basin of the critical point $({1\mathord{\left/ {\vphantom {1 a_{*} }} \right. \kern-\nulldelimiterspace} a_{*} } )^{1/2} $, and later return to the attraction basin of the critical point $0$. The other real solution does not originate a breaking point because it does not verify Eq.~\eqref{ec:case2conditions}.

Equation~\eqref{ec:case3-1} gives a solution that it is not a breaking point because is out of the interval [$a_{1} $, $a_{3} $].

A solution of Eq.~\eqref{ec:case3-2} is $a_{5} =1.099012$, Fig.~\ref{fig:6}(c). When $a>a_{5} $, the orbit is in the attraction basin of the critical point $0$ and it reaches a fixed point. When $a<a_{5} $, the orbit is in the attraction basin of the critical point $({1\mathord{\left/ {\vphantom {1 a_{*} }} \right. \kern-\nulldelimiterspace} a_{*} } )^{1/2} $ and it is chaotic. The solution $a_{5} $ is a breaking point, Fig.~\ref{fig:4}(a,c). The other solution of Eq.~\eqref{ec:case3-2} is out of the working intervals of systems A and B. The fixed point $x_{{\rm 1}} =-0.076534$, Fig.~\ref{fig:4}(a) and Fig.~\ref{fig:6}(c), is the solution $-1<x_{{\rm 1}} <0$ of the equation $1-a_{5} (1-a_{*} x^{2} )^{2} =x$ and the fixed point $x_{{\rm 2}} =0.989720$, Fig.~\ref{fig:4}(c), is the solution $0.9<x_{{\rm 2}} <1$ of the equation $1-a_{*} (1-a_{5} x^{2} )^{2} =x$.

\begin{figure}[h]
\begin{center}
\includegraphics[width=85mm]{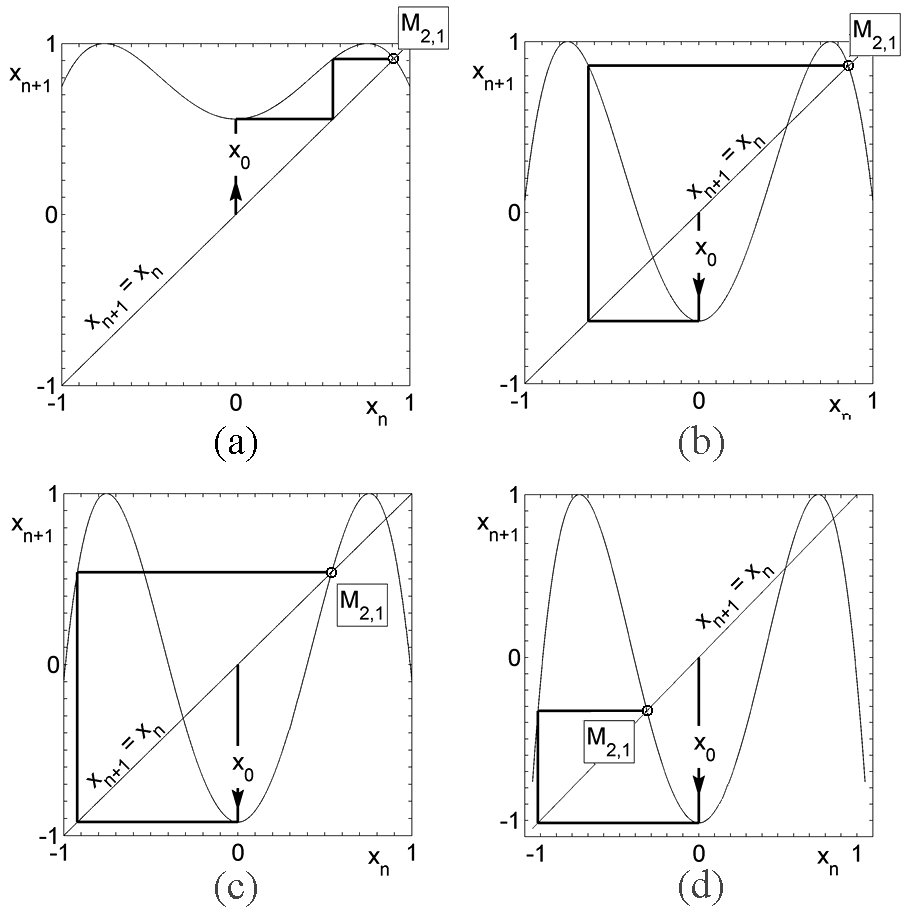}\\
\caption{Graphical iteration of quartic $x_{n+1}=1-a(1-a_{*}x_n^2)^2$ when $a_*=1.754677$. (a) $a=0.441338$. (b) $a=1.634264$. (c) $a=1.921352$. (d) $a=2.016013$.} \label{fig:7}
\end{center}
\end{figure}

Equation~\eqref{ec:tensolutionfora} has four real solutions for $a$ that give rise to Misiurewicz points ${\rm M}_{{\rm 2,1}} $, Fig.~\ref{fig:7}. According to what we have said, none of these solutions creates a breaking point.\\

\subsection{Bifurcation diagrams when $a=1.381547$}

\begin{figure*}[!]
\begin{center}
\includegraphics[width=165mm]{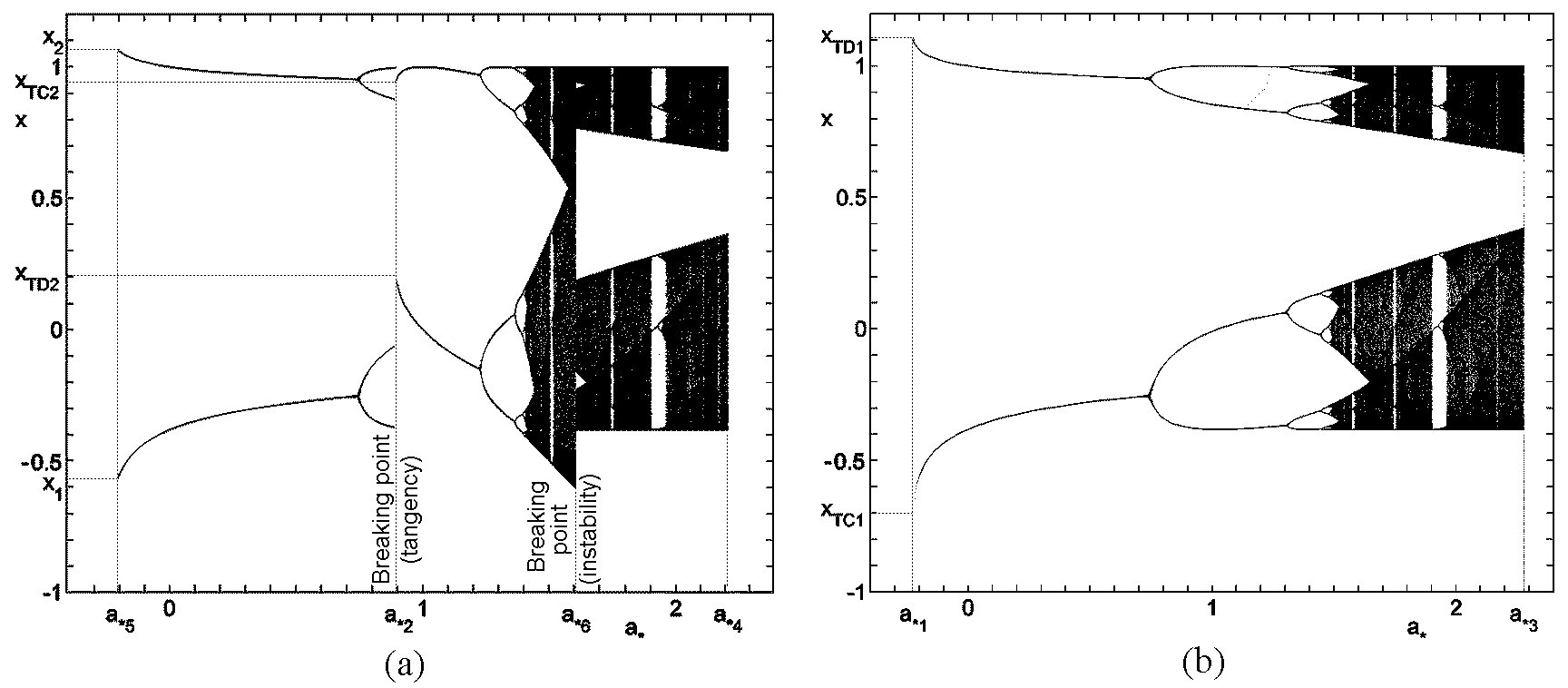}\\
\caption{(a) Bifurcation diagrams of systems A and B when $a=1.381547$. (a) System A. (b) System B.} \label{fig:8}
\end{center}
\end{figure*}

The map $x_{n+1} =1-ax_{n}^{2} $ has a superstable period-8 orbit when $a=1.381547$. Let us consider, as an example, the bifurcation diagrams of systems A and B when $a=1.381547$ (Fig.~\ref{fig:8}).

\subsubsection{Breaking point by tangency}

Equation~\eqref{ec:quarticsD1} gives $x_{{\rm TD1}}=1.117409$ and $x_{{\rm TD2}} =0.215923$. Equation~\eqref{ec:quarticsD2} gives $a_{*1} =\!\!-0.223368$ and $a_{*2} =0.895757$. The point $a_{*1} $, the lower extreme of the working interval of system B, is not a breaking point, Fig.~\ref{fig:8}(b). The point $a_{*2} $ is a breaking point, Fig.~\ref{fig:8}(a). When $a_{*} =a_{*1} $, Eq.~\eqref{ec:tangentpoint} gives $x_{{\rm TC1}} =\!-0.725005$ that is a valid solution because Eq.~\eqref{ec:2tangentpointparam} restores $a$, Fig.~\ref{fig:8}(b). Similarly, when $a_{*} =a_{*2} $, Eq.~\eqref{ec:tangentpoint} gives $x_{{\rm TC2}} =0.935587$ that is a valid solution because Eq.~\eqref{ec:2tangentpointparam} restores $a$, Fig.~\ref{fig:8}(a).

\subsubsection{Breaking points by instability}

A solution of Eq.~\eqref{ec:2solutionforaasterisco} is $a_{*3} =2.273223$ that originates a Misiurewicz point ${\rm M}_{{\rm 2,1}} $, Fig.~\ref{fig:8}(b). Note that Eq.~\eqref{ec:2solutionforaasterisco} is obtained starting from the quartic C with $x_{0} =0$, i.e., for system B. Therefore, the solution $a_{*3} $ is the upper extreme of the working interval of system B and it is not a breaking point. The other solution is out the working interval of the bifurcation diagram.

Equation~\eqref{ec:case3posibleBP} gives $a_{*4} =2.203184$ and $a_{*5} =\!-0.203184$, Fig.~\ref{fig:8}(a), the extremes of the working interval of the bifurcation diagram of system A. They are not breaking points according to Eq.~\eqref{ec:case3BPcondition}.
 The fixed point $x_{1} =-0.569911$ is a solution of $1-a(1-a_{*}^{5} x^{2})^{2}= x$  and the fixed point $x_{2} =1.065993$ is a solution of $1-a_{*5} (1-ax^{2} )^{2} =x$, Fig.~\ref{fig:8}(a).


Equation~\eqref{ec:case3realsolutions} has two real solutions. The solution $a_{*6} =1.603002$,  Fig.~\ref{fig:8}(a), verifies Eq.~\eqref{ec:case3BPcondition} and it is a breaking point. In the quartic C, when $x_0=(1/a_{*6})^{1/2}$, if $a_{*} $ is slightly smaller than $a_{*6} $, the orbit is chaotic inside the attraction basin of the critical point $(1/a_{*6})^{1/2}$. If $a_{*} $ is slightly greater than $a_{*6} $, the orbit is chaotic inside the attraction basin of the critical point  $0$. The other solution does not verify Eq.~\eqref{ec:case3BPcondition}.\\


\section{Conclusions}\label{sec:conclusion}
The bifurcation diagram of the alternate iteration of the quadratic map $1-ax^{2} $ is double, as a consequence of Property 1, because it is constituted by the bifurcation diagrams of two quartic maps that operate simultaneously.

The bifurcation diagrams of these quartic maps present breaking points, i.e. parameter values where the bifurcation diagrams change abruptly from chaos to period-1 or from a band of chaos to another band of chaos. Therefore, the bifurcation diagram of the alternate quadratic map $1-ax^{2} $ also presents breaking points. Equations to calculate the parameter values corresponding to these points are deduced, and examples are shown.

\textbf{Acknowledgments} \noindent This work has been partially supported by Ministerio de Ciencia e Innovaci\'{o}n (Spain) under the Grant no. TIN2011-22668.

\bibliographystyle{apalike}

\end{document}